%% file: stab-stt-4c.tex
\documentclass[11pt,fleqn]{article}
\usepackage{bezier,amsmath,amsfonts}
\usepackage[dvips]{graphics}
\usepackage{hyperref}
\usepackage{float}

\input preamble.tex

\def\eqn#1{\eq\eqref{#1}}
\def\rf{\eqref}
\def\besub{\begin{subequations}}
\def\esub{\end{subequations}}

\def\M{{\mathbb M}}
\def\ME{\mbox{$\M_{\rm E}$}}
\def\MJ{\mbox{$\M_{\rm J}$}}

\def\og{{\overline g}}

\def\oR{{\overline R}}

\def\df{\delta\psi}
\def\da{\delta\alpha}
\def\db{\delta\beta}
\def\dg{\delta\gamma}

\def\cF{{\cal F}}

\def\cL{{\cal L}}

\def\tT{\widetilde{T}}
\def\kappa{\varkappa}

\def\Veff{V_{\rm eff}}

\def\GR{general relativity}
\def\pb{perturbation}
\def\pbs{perturbations}
\def\sph{spherically symmetric}
\def\ssph{static, spherically symmetric}

\def\wh{wormhole}
\def\whs{wormholes}
\def\bh{black hole}
\def\bhs{black holes}

\def\emag{electromagnetic}
\def\grav{gravitational}

\def\Schr{Schr\"o\-din\-ger}
\def\RN{Reiss\-ner-Nord\-str\"om}
\def\BD{Brans-Dicke}

\def\mn{_{\mu\nu}}
\def\MN{^{\mu\nu}}
\def\mN{_\mu^\nu}

\usepackage{color}

\allowdisplaybreaks
\begin{document}
\onecol

\Title{Nonlinear electrodynamics and stability of spherically symmetric\yy
	 	space-times in scalar-tensor gravity}

\Aunames{Kirill A. Bronnikov,\au{a,b,c,1}
		Rustam Ibadov,\au{d,2}, Feruza Y. Shaymanova,\au{d,f,3}\\ 
		Najibullokhon M. Shukurullokhon\au{d,4}}
	
\Addresses{\small
\adr a	{Center of Gravitation and Fundamental Metrology, VNIIMS, 
		Ozyornaya ul. 46, Moscow 119361, Russia}
\adr b	{Institute of Gravitation and Cosmology, RUDN University, 
		ul. Miklukho-Maklaya 6, Moscow 117198, Russia}
\adr c 	{National Research Nuclear University ``MEPhI'', 
		Kashirskoe sh. 31, Moscow 115409, Russia}
\adr d {Department of Theoretical Physics and Quantum Electronics, 
		Institute of Engineering Physics,\\ Samarkand State University 
		named after Sharof Rashidov, Samarkand, 140104, Uzbekistan}
\adr e {Karshi State  University, Kochabog street, Karshi city 180103, Uzbekistan}
		}

\Abstract
  {We study linear perturbations of static, spherically symmetric solutions of  
   scalar-tensor theories (STT) of gravity from the Bergmann-Wagoner-Nordtvedt class,
   sourced by nonlinear electrodynamics (NED). We obtain a general expression for
   the effective potential $\Veff$ governing the perturbation dynamics for theories with 
   arbitrary scalar-electromagnetic interaction of the form $L(\psi, F)$, where $\psi$ is a 
   scalar field and $F = F\mn F\MN$ the electromagnetic invariant. This consideration 
   includes, in particular, arbitrary scalar self-interaction potentials and scalar fields that 
   can be phantom in some regions of space-time (the so-called trapped ghosts). Only radial
   (monopole) perturbations are considered here as the most likely ones to cause an instability. 
   It is shown, in particular, that if NED has a correct Maxwell weak field limit, the zero charge 
   limit of $\Veff$ does not contain any trace of NED, and the perturbation dynamics is the 
   same as for vacuum STT solutions. The previously obtained stability results for 
   STT-Maxwell solutions are shown to be extended without change to STT-NED solutions
   with equal electric and magnetic charges, implying $F =0$. 
    }

\email 1 {kb20@yandex.ru} 
\email 2 {ibrustam@mail.ru}
\email 3 {yusupovnafiruz89@gmail.com}
\email 4 {Najibulloxon@samdu.uz}

\section{Introduction}
  
  The advent of nonlinear electrodynamics (NED) was largely motivated by the prospect of infinity
  avoidance in \emag\ self-energy of electric point charges (M. Born and L. Infeld, \cite{born34}) 
  and by invoking quantum effects of electron-positron interaction as corrections to Maxwell's
  theory (W. Heisenberg and H. Euler, \cite{heis36}). 
  Much later, other forms of NED were incorporated into gravitational frameworks aiming at
  resolving space-time singularities and regular black hole construction \cite{pell69, kb01, fan-wang}. 
  Furthermore, it was discovered that Born-Infeld-like Lagrangian structures naturally 
  emerge in the low-energy effective limits of superstring theory and quantum electrodynamics 
  with one-loop corrections \cite{frad85, sorokin22}. In parallel, scalar-tensor theories (STT),
  beginning with the \BD\ theory \cite{BD-STT} and the more general 
  Bergmann-Wagoner-Nordtvedt (BWN) class \cite{STT1, STT2, STT3} and ending with wider
  Horndeski \cite{horn} and beyond-Horndeski \cite{beyond-horn} classes of theories 
  provide a natural means of extending general relativity (GR) to address cosmological 
  phenomena  and strong-field gravity regimes \cite{quiros19}. Despite the extensive literature 
  on static background configurations within isolated NED or STT models and their stability
  properties, a deep study of their combined dynamical behavior remains fundamentally 
  challenging.

  Such studies should naturally begin with a search for and analysis of self-gravitating \ssph\
  configurations described by scalar-electrovacuum solution generalizing the known scalar-Maxwell 
  ones \cite{penney, br73, fara21}. The simplest examples of such solutions are those belonging to
  so-called dilaton gravity, that is, GR with matter in the form of a scalar field $\phi$ minimally 
  coupled to gravity but interacting with the Maxwell field via the term $\e^{2\lambda\phi} \cF$,
  where $\cF = F\mn F\MN$ and $\lambda = \const$ \cite{shik77, gibb88, garf91}. These 4D
  solutions were later extended to systems with extra dimensions, multiple antisymmetric 
  forms as generalizations of the Maxwell field and multiple scalar fields, being largely
  motivated by developments in supergravity and superstring theories, see, e.g., \cite{im01} 
  for a review. Such solutions generically contain naked singularities, but there are also families 
  of \bh\ solutions. The simplest representatives of these solutions were subject to stability
  studies  \cite{bm95, bm00}, and the main conclusion was that solutions with naked singularities 
  generically suffer a catastrophic instability under \sph\ \pbs, quite similarly to the instability 
  found in \cite{kb-hod} for Einstein-scalar and Einstein-Maxwell-scalar solutions with 
  naked singularities, while \bhs\ form a kind of stability islands \cite{bm95, bm00}. 
  
  A number of solutions have been obtained and analyzed for the system of NED coupled to 
  a dilatonic scalar field in the framework of GR or STT using numerical methods
  \cite{cle00, torii00, yaz1, yaz2, yaz3}, mostly for the Born-Infeld version of NED. 
  The corresponding sets of equations turn out to be too complicated for obtaining analytical 
  solutions; still some particular examples of such solutions have been found in \cite{riazi06}
  for the dilaton-Born-Infeld system in GR with or without Liouville-type scalar field potentials. 
  
  More complex exact solutions involving scalar fields and NED can be obtained using the 
  result of \cite{rahul} that any \ssph\ metric can be obtained as a solution of GR with a
  magnetic or electric NED source with a Lagrangian $\cL(\cF)$ plus a scalar field $\phi$ with 
  a nonzero potential $V(\phi)$. Examples of such solutions can be found in 
  \cite{rahul, trap22, kb24a, kb24b}; their common shortcoming consists in rather cumbersome  
  expressions for $\cL(\cF)$ containing fractional powers of $\cF$.
  
  As to stability studies, such an analysis have been carried out in \cite{yaz4} for \sph\ \pbs\ 
  of a family of scalar-tensor \bhs\ with Born-Infeld NED, and it was concluded that such 
  objects are linearly stable due to a nonnegative effective potential for these \pbs. This 
  confirms the observation of \cite{bm95, bm00} that \bh\ solutions, which are special with 
  respect to the whole set of solutions involving scalar fields, form a kind of stability islands. 
  A more general conclusion on \bh\ stability was obtained in \cite{soda20}: it concerned 
  magnetic black holes in GR sourced by a general class of NED with Lagrangians given as 
  arbitrary functions of $F\mn$ and its Hodge dual $^*F\mn$ and linear \pbs\ of any 
  multipolarity.
   
  It is, however, worthwhile to mention another important result that encompasses regular 
  \bhs\ or horizonless solitonic objects sourced by NED with or without scalar fields 
  of sufficiently general nature (e.g., including k-essence versions of the scalar) \cite{tsuj24}:
  it has been shown that near a regular center vector \pbs\ always exhibit an angular   
  Laplacian instability. In other words, angular components of such vector \pbs\ obey equations
  implying that their propagation speed is imaginary, leading to their exponential growth 
  in time. This rules out such regular black holes and solitons as possible steady objects. 
  Combined, the results of \cite{soda20} and \cite{tsuj24} can mean that one may deal with 
  regular \bhs\ whose domain of outer communication is stable (only this domain is generally 
  subject to stability analysis) but there is a fundamentally unstable region in their deep interior. 
  
  Let us note that non-\bh\ solutions, including those with naked singularities, also deserve a
  careful stability analysis since, on one hand, in the presence of scalar fields such solutions are
  more general than \bh\ ones, and on the other, there is a common hope that quantum \grav\ 
  effects should regularize all singularities of classical field systems, whose stability properties
  may be extended to such smoothed systems. As long as the stability analysis involves 
  solving boundary-value problems for \pbs, one should require that the boundary conditions 
  for \pbs\ of classical singular systems should be formulated so as to survive their possible 
  regularization.
  
  One should also recall that the BWN class of STT admits scalar field-dependent conformal 
  mappings connecting Jordan, Einstein and, in general, other conformal frames. In the field 
  equations, such mappings reduce to mere substitutions of variables, hence it is possible 
  to apply the same equations to describe the field dynamics in any frame. It is often convenient
  to employ the Einstein frame that coincides with GR with the corresponding material 
  sources. Meanwhile, the boundary conditions for \pb\ analysis should be formulated in the 
  physically preferred frame, which is usually Jordan's, distinguished by the condition of 
  minimal coupling between matter and gravity. The conformal factor between the Einstein and 
  Jordan frames depends on the particular choice of STT, therefore, the final stability inferences 
  for similar field systems can be different in different STT. 
  
  The stability of \ssph\ vacuum and electrovacuum space-times in the BWN class of STT under 
  \sph\ \pbs\ was studied in \cite{we-1, we-2}, extending the results obtained previously 
  in the framework of GR \cite{kb-hod}.  The study reduced to solving boundary-value 
  problems for a unique master equation,  determined in the common Einstein frame of all
  these STT, but, as said above, due to different boundary conditions, the inferences turned 
  out to be different for different particular STT as well as the solution parameters. As examples, 
  we considered the \BD\ \cite{BD-STT}, Barker \cite{barker} and Schwinger \cite{schwg} STT 
  and also GR with nonminimally coupled scalar fields with the term $\xi R \phi^2$ in the 
  Lagrangian that can also be treated as STT cases. 
  
  In the present paper, we try to extend the study to nonlinear \emag\ fields. As a tool for 
  such analysis, we consider a sufficiently general scalar-\emag\ interaction theory (see \eqn{S}) 
  in GR (hence in the Einstein frame of BWN STT) and obtain a single master equation for \pb\ 
  modes with a certain effective potential $\Veff$ that, along with physically relevant 
  boundary conditions, should determine the \pb\ dynamics.  
  
  For the first application of this general potential, we make use of the following circumstance: 
  in the class of NED theories described by Lagrangian functions $\cL(\cF)$ with a correct 
  Maxwell-like weak field behavior, there always exist solutions with equal electric and magnetic 
  charges such that $\cF=0$ \cite{russo26,dyon-we}, and in this case the \emag\ field exhibits 
  a Maxwell-like behavior. It follows, in particular, that any known electrovacuum solution 
  of STT, where the \emag\ field obeys the Maxwell equations, may be re-interpreted as a 
  NED-STT solution. Meanwhile, \pbs\ of such systems may in principle feel the influence of a 
  particular NED theory, which should then affect the stability of the corresponding static solution.  
  However, analyzing the expression for $\Veff$, we come to the conclusion that the \pb\ 
  dynamics of such background solutions remains the same as was previously obtained \cite{we-2} 
  for Maxwell-STT solutions. We also found no effect of using NED instead of Maxwell's theory 
  on the stability of the family of exceptional Brans-Dicke \whs\ studied in \cite{we-25}.
  
  The paper is organized as follows. In Section 2 we consider the BWN STT-NED equations in 
  Jordan and Einstein frames for an arbitrary NED with $\cL = \cL(\cF)$ and obtain the Einstein 
  frame in terms of the accordingly modified version $F$ of the \emag\ invariant. In Section 3, 
  we derive the \Schr-like equation for \sph\ \pbs\ in a more general NED-scalar system 
  with the Lagrangian $L = L(\psi, F)$ in GR or, equivalently, the Einstein frame of STT. 
  In Section 4 we briefly describe the STT-Maxwell and special STT-NED solutions, their 
  stability properties are discussed in Section 5, and Section 6 is a conclusion.   
  
\section{NED-STT equations in Jordan and Einstein frames}

  Consider the action of the general Bergmann-Wagoner-Nordtvedt STT of gravity 
  \cite{STT1, STT2, STT3} coupled to NED with the Lagrangian functions $\cL(\cF)$, 
\bearr   \label{S_J}
             S_{\rm STT} = \frac 1{16\pi} \int \sqrt{-g} d^4 x
		             \Big[f(\phi) R + 2 H(\phi)g\MN\phi_{,\mu} \phi_{,\nu} 
  			           - 2 U(\phi) -\cL(\cF)\Big],
\ear    
  where $R$ is the scalar curvature, $g = \det(g\mn)$, $f(\phi), H(\phi)$, and $U(\phi)$ 
  are arbitrary functions ($f(\phi) > 0$ describes a nonminimal coupling between the 
  space-time curvature and the scalar field $\phi$); furthermore, $\cL(\cF)$ is an arbitrary 
  smooth function of the \emag\ invariant $\cF = F\mn F\MN$, 
  $F\mn = \D_\mu A_\nu -  \D_\nu A_\mu$, and we assume that $\cL(\cF)$ has a correct 
  Maxwell limit at small $\cF$, that is, $\cL(\cF) \approx \cF$ as $\cF \to 0$. 

  The formulation \rf{S_J} of STT corresponds to Jordan's (conformal) frame, specified in 
  pseudo-Riemannian space-time \MJ\ with the metric $g\mn$; it is physically distinguished 
  by the postulate that matter (including NED) is there minimally coupled to gravity, so that,
  for example, neutral test particles follow geodesic paths.
  
  As is widely known \cite{STT2}, there is a conformal mapping that converts theories with 
  the action like \rf{S_J} to a form inherent to \GR\ but in general with a nonminimal coupling
  between matter and gravity, called the Einstein conformal frame. The theory is then specified 
  in space-time \ME\ with the metric
\beq              \label{map}
		\og\mn = f(\phi)\, g\mn.
\eeq  
  The action \rf{S_J} transformed to \ME\ takes the form 
\bearr   \label{S_E}
             S_{\rm STT} = \frac 1{16\pi} \int \sqrt{-\og} d^4 x
		             \bigg[\oR + 2 \eps \og\MN \psi_{,\mu} \psi_{,\nu} 
  			           - 2 \frac {U(\phi)}{f^2(\phi)} + L(\psi, F)\bigg],
\ear      
  where overbars mark quantities obtained from or with $\og\mn$, and it is meant that the field
  $\phi$ has been expressed in terms of the new field $\psi$, minimally coupled to the 
  metric $\og\mn$, according to the relations
\beq  			\label{phi-psi}
		\frac {d\phi}{d\psi} = \frac{\sqrt 2 f(\phi)}{\sqrt{|D|}}, \qq
		D = f(\phi)H(\phi) + \frac 32 \Big(\frac{df}{d\phi}\Big)^2, \qq \eps = \sign D.
\eeq  
  Thus, $\eps =1$ corresponds to a canonical nature of the field $\psi$ (and hence $\phi$), while 
  in the case $\eps =-1$, the scalar $\psi$ (and hence $\phi$) has negative kinetic energy and 
  is called a ghost or phantom field. As to NED, the map \rf{map} converts the Lagrangian 
  $\cL(\cF)$ to 
\beq                \label{Lpsi}
		L(\psi, F) =  \frac 1{f^2(\phi)} \cL(\cF) = \frac 1{f^2(\phi(\psi))} \cL\big(f^2(\phi(\psi)) F\big),
\eeq  
  where $\cF$ and $F$ are the \emag\ invariants calculated in \MJ\ and \ME, respectively:
\beq
		\cF = g^{\mu\alpha}g^{\nu\beta} F\mn F_{\alpha\beta}, \qq
		  F = \og^{\mu\alpha}\og^{\nu\beta} F\mn F_{\alpha\beta}, \qq
		  g\MN = f \og\MN, \qq
		  \cF = f^2 F.
\eeq   
  
  Evidently, if we know a solution to the field equations in $\ME$, it is easy to obtain its counterpart 
  in $\MJ$ using the transformation \rf{map} and \rf{phi-psi}, so that the line element in \MJ\ is
\beq             \label{ds_J}
		  ds_J^2 = g\mn dx^\mu dx^\nu = \frac 1{f(\phi(\psi))} ds_E^2,
\eeq   
  and the scalar and matter variables are reformulated accordingly. This concerns any kind of 
  solutions, irrespective of the space-time symmetry and specific coordinate dependences. Of 
  interest for us here will be \ssph\ solutions and their small \pbs\ preserving spherical symmetry.
        
\section{Perturbations of NED-STT solutions: the Einstein frame}

  When trying to answer the question posed in the introduction on the stability of static 
  NED-STT solutions, it makes sense to use their representation in \ME, in which case they obey 
  the equations due to \rf{S_E} that are easier for solution and analysis. Also, bearing in 
  mind possible employment of the results of this analysis in other related problems, we will
  start with an action more general than \rf{S_E}: 
\beq                                                  \label{S}
     {\cal S} = \frac 1 {16\pi} \int {\sqrt{-g}}
	\bigg(R + 2H(\psi) g^{\alpha\beta}\psi_{,\alpha}\psi_{,\beta} - L(\psi, F)\bigg),
\eeq
  where the ``kinetic function'' $H(\psi)$ is positive for a canonical scalar field with positive 
  energy density, and $H(\psi)<0$ for a phantom scalar field. There are also cases of interest where 
  the function $H(\psi)$ smoothly changes its sign, being negative, say, in a strong field region 
  and positive outside it (the so-called ``trapped ghost'' \cite{trap10, trap17, stab18, trap22}); 
  we will not exclude here this opportunity. As before, $F\mn$ is the electromagnetic field tensor,
  $F = F\mn F\MN$, and we consider the function $L(\psi, F)$ to be quite arbitrary; the function 
  \rf{Lpsi} is its special case, but in general it may also include a contribution of a scalar 
  field potential $U(\phi)$ as given in \rf{S_J} or \rf{S_E}. The units are chosen so that 
  $c = 8\pi G = 1$. Our aim is to obtain the effective potential in the \Schr-like equation for 
  modes of time-dependent spherical \pbs\ of a \ssph\ solution, the latter assumed to be known. 
  Such \sph\ \pbs\ contain a single dynamical degree of freedom related to the scalar field 
  \pb\ $\df$ and are much easier to analyze than more general \pbs\ with multipolarities 
  $\ell > 0$, involving vector and tensor modes; on the other hand, as follows from rich 
  experience, \sph\ \pbs\ are the most ``dangerous'' ones from the viewpoint of instability since 
  their effective potentials do not contain the centrifugal term $\ell(\ell +1)/r^2$ that can only
  increase the eigenvalues as compared to $\ell=0$ and thus weaken or even eliminate
  a possible instability.
  
  In this section, we consider $g\mn$ as the Einstein-frame metric with no risk of confusion.

\paragraph{Basic equations.} The field equations following from \rf{S} have the form
\bearr
    	4 h \nabla^\mu \nabla_\mu \psi
    		+ 2H_\psi \psi_{,\mu}\psi^{,\mu} + L_\psi = 0,        	   \label{Ephi}
\\  \lal
        \nabla_\mu (L_F\, F\MN) = 0,		   \label{Eemag}
\\    \lal       \label{EE}
       	G\mN \equiv R\mN - \half \delta\mN R = - T\mN, \ \ {\rm or\ equivalently}\ \  
       	R\mN = - \tT\mN := - (T\mN - \half \delta\mN T^\alpha_\alpha)	  	   
\ear
  where the indices $\psi$ and $F$ denotes derivatives with respect to $\psi$ and $F$, 
  respectively, and $T\mN = T\mN [\psi] + T\mN [F]$ is the stress-energy tensor (SET), where
\bearr
      T\mN [\psi] = H(\psi)[ 2\psi_\mu \psi^\nu - \delta\mN \psi^{\alpha}\psi_{\alpha}],
\nnn
      T\mN [F] = -2 L_F F_{\mu\alpha}F^{\nu\alpha}+ \half \delta\mN L(F, \psi).  \label{SET}
\ear

  The general \sph\ metric $ds^2 = g\mn dx^\mu dx^\nu$ may be written in the form
 \beq                                                         \label{ds-E}
    ds^2 = \e^{2\gamma(u,t)} dt^2 - \e^{2\alpha(u,t)}du^2 - \e^{2\beta(u,t)}d\Omega^2,
			       \qq d\Omega^2:= d\theta^2 + \sin^2\theta d\varphi^2,
\eeq    
  we also use the notation $r(u,t) \equiv \e^\beta$.
  A center (if any) corresponds to $r \to 0$ under the condition $\e^{2\gamma} \geq 0$.

  We consider linear \sph\ perturbations of static solutions to the field equations due to 
  (\ref{S}), which we assume to be known. Accordingly, the metric has the form
  (\ref{ds-E}) but now $\gamma(u,t) = \gamma(u) + \delta\gamma(u,t)$ and similarly
  for other quantities, in particular, $\psi (u,t) = \psi(u) + \delta\psi(u,t)$. 
  The most general electromagnetic field $F\mn$ compatible with spherical symmetry 
  consists of radial electric and magnetic fields with the only nonzero components
  $F_{01} = - F_{10}$, $F_{23} = - F_{32}$, and we have
\bearr                                                          \label{F10}
	L_F \e^{\alpha+2\beta+\gamma} F^{10} = q_e,
\\ \lal                                                \label{F}
	F_{01} = - F_{10} = \frac{q_e}{L_F} \e^{\alpha - 2\beta +\gamma},
	\qq 
	F_{23} = - F_{32} = q_m \sin\theta, 
	\qq
	F = 2\e^{-4\beta}\bigg(q_m^2 -  \frac {q_e^2}{L_F^2}\bigg) ,
\ear

  It will be convenient to use the Einstein equations \rf{EE} as $R\mN = \ldots$. To that end, 
  preserving only linear terms with respect to time derivatives, we can write all 
  nonzero components of the Ricci tensor as
\bear
     R^0_0 \eql                                         \label{R00}
     \e^{-2\gamma}(\ddot\alpha + 2\ddot\beta)
           -\e^{-2\alpha}[\gamma'' +\gamma'(\gamma'-\alpha'+2\beta')],
\yy
     R^1_1 \eql
     \e^{-2\gamma}\ddot\alpha                           \label{R11}
     - \e^{-2\alpha}[\gamma''+2\beta'' +\gamma'{}^2+2\beta'{}^2
            -\alpha'(\gamma'+2\beta')],
\yy
     R^2_2 \eql R^3_3 = \e^{-2\beta}                                    \label{R22}
          +\e^{-2\gamma}\ddot\beta
              -\e^{-2\alpha}[\beta''+\beta'(\gamma'-\alpha'+2\beta')],
\yy
     R_{01}\eql
        2[\dot\beta' + \dot{\beta}\beta'                        \label{R01}
                 -\dot{\alpha}\beta'-\dot{\beta}\gamma'],
\ear
  where dots and primes denote $\D/\D t$ and $\D/\D u$, respectively.
  In the right-hand sides of \eqs\rf{EE}, the only nonzero components 
  of the tensors $\tT\mN[\psi]$ and $\tT\mN[F]$ read in the same approximation
\bearr     \label{T-phi}
		\tT^1_1[\psi] = -2 \e^{-2\alpha}H(\psi)\psi'^2,  \qq
		\tT_{01}[\psi] = \tT_{10}[\psi] = 2 H(\psi) \psi' \dot \psi,
\nnn      \label{T-F0}
		\tT^0_0[F] = \tT^1_1[F ] =- \half L + 2q_m^2 L_F \e^{-4\beta},		
\nnn	    \label{T-F2}	
		\tT^2_2[F] = \tT^3_3[F ] =- \half L - 2q_e^2 L_F^{-1} \e^{-4\beta}.	
\ear  

  The zero-order (i.e., static) scalar field equation and the $({}^0_0)$, $({}^1_1)$, $({}^2_2)$
  components of \eqs (\ref{EE}) are
\bearr                                                      \label{back}
     \psi'' + \psi'(\gamma'+2\beta'-\alpha')
     		+ \frac {H'}{2H} \psi' - \frac 1{4H}\e^{2\alpha}L_\psi =0,
\nnn                                                        
     \gamma'' + \gamma'(\gamma'+2\beta'-\alpha') = 
     						- \e^{2\alpha} \Big(\half L - 2q_m^2 L_F \e^{-4\beta}\Big),
\nnn                                                        
     \gamma'' + 2\beta'' + \gamma'{}^2 + 2\beta'{}^2 -\alpha'(\gamma'+2\beta') 
     		=  - 2H \psi'^2 - \e^{2\alpha} \Big(\half L - 2q_m^2 L_F \e^{-4\beta}\Big),		
\nnn                                                       
     \e^{2\alpha-2\beta}
       - \beta'' - \beta'(\gamma'+2\beta'-\alpha') 
       =  \e^{2\alpha}\Big( \half L + 2q_e^2 L_F^{-1} \e^{-4\beta}\Big).
\ear
  The first-order \pb\ equations (scalar, $R_{01}=\ldots$, and $R^2_2 = \ldots$) read
\bearr                                                     \label{e-phi1}
      -\e^{2\alpha-2\gamma} \delta\ddot\psi 
      + \df''  + \df' (\gamma'+2\beta'-\alpha') +\psi'(\dg' + 2\db' -\da')]
\nnn \inch\inch
	 + \frac{H'}{H} \df' + \Half \psi'^2 \delta\Big(\frac{H_\psi}{H}\Big) 
            - \frac 14 \delta\Big(\frac{L_\psi \e^{2\alpha}}{H}\Big) = 0,
\yyy                                                       \label{01-1}
    \delta\dot\beta' + \beta'\delta\dot\beta
        - \beta' \delta\dot\alpha - \gamma' \delta\dot{\beta}
                = - H \psi'\delta\dot\psi,
\yyy                                                       \label{22-1}
    \delta(\e^{2\alpha-2\beta})
            + \e^{2\alpha-2\gamma} \delta\ddot\beta
                    -\db'' -\db'(\gamma'+2\beta'-\alpha')  -\beta'(\dg' + 2\db' -\da')  
\nnn \inch \inch                   
         = \Half \delta(L \e^{2\alpha}) + 2q_e^2 \delta(L_F^{-1}\e^{2\alpha - 4\beta}).
\ear
  Equation \rf{01-1} is easily integrated in $t$; since we are interested in
  time-dependent perturbations, we omit the emerging arbitrary function of $u$ that 
  describes static perturbations and obtain
\beq                                                       \label{01-1i}
    \db' + \db(\beta'-\gamma') - \beta'\da = -H \psi' \df.
\eeq

  Let us note that we have two independent forms of arbitrariness: one is the
  freedom of choosing a {\it radial coordinate\/} $u$, the other is a {\it
  perturbation gauge\/}, or, in other words, a reference frame in the
  perturbed space-time, which can be expressed in imposing a certain
  relation for $\da,\ \db$, etc. All the above equations have been written in the most
  universal form, without coordinate or gauge fixing.

\paragraph{Gauge $\db \equiv 0$.}
  This gauge is technically the simplest one, it only causes certain problems
  when applied to configurations with extremum values of $r$ (such as \wh\ throats) since 
  it leaves invariable, e.g., the throat radius, while perturbation must in general admit its 
  time dependence \cite{gon08,stab11,stab12}. This difficulty is irrelevant in the present study.

  With $\db = 0$, \eq (\ref{01-1i}) expresses $\da$ in terms of $\df$:
\beq                                  			      \label{da}
       \da = \frac {H(\psi) \psi' }{\beta'} \df.
\eeq
  Furthermore, \eq (\ref{22-1}) expresses $\dg' - \da'$ in terms of $\da$ and $\df$:
\beq                                  \label{dg-da}
    \beta'(\dg'-\da') = 2 \e^{2\alpha-2\beta}\da
                - \Half \delta(L \e^{2\alpha}) -2 q_e^2 \e^{-4 \beta} \delta(L_F^{-1}\e^{2\alpha}).
\eeq
   Our aim is to express in terms of $\df$ \pbs\ of all other quantities and then substitute 
   them to \eq \rf{e-phi1}. In particular, for $L(\psi,F)$ we have 
   $\delta L = L_F \delta F + L_\psi \df$, while for $\delta F$ we obtain from \rf{F}
\beq      \label{dF0}
		\delta F = -2 q_e^2 \e^{-4\beta} \delta (L_F^{-2})
				= 4 q_e^2 \e^{-4\beta} L_F^{-3} (L_{FF}\delta F + L_{F\psi}\df).
\eeq      
   From this linear algebraic equation for $\delta F$ we obtain 
\beq      \label{dF}
	\delta F  = \frac{4q_e^2 L_{F\psi}}{L_F^3 \e^{4\beta} - 4q_e^2 L_{FF}}\df \equiv P(u) \df. 
\eeq   
   The so defined coefficient $P = P(u)$ is a combination of quantities belonging to the static 
   background solution. Some care is necessary due to its possible blowing up when and if 
   the denominator turns to zero. 

  Substituting all that into (\ref{e-phi1}), using \rf{dg-da} for $\dg'-\da'$, \rf {da} for $\da$ 
  and \rf{dF} for $\delta F$, we finally obtain the following wave equation:
\bearr                                                        \label{eq-df}
      \e^{2\alpha-2\gamma} \delta\ddot\psi
            -\df'' - \df' (\gamma'+ 2\beta'-\alpha'+ H'/H) + W(u) \df =0,
\yyy                                    \label{W}
      W(u) \equiv \frac{2 H \psi'^2}{\beta'^2} \e^{2\alpha}
      			\bigg( \frac L2 - \e^{-2\beta} + 2q_e^2\frac{\e^{-4\beta}}{L_F} \bigg)
	+ \frac{\psi'}{\beta'}\e^{2\alpha}
	\bigg[L_\psi +\Half P L_F - 2q_e^2 \e^{-4\beta}\frac{L_{F\psi} +P L_{FF}}{L_F^2} \bigg] 
\nnn \inch	
	+ \frac 14 \e^{2\alpha}\bigg[\frac 1H(L_{\psi\psi} + P L_{F\psi}) 
				- \frac {H_\psi}{H^2} L_\psi\bigg]   
				+ \Half \psi'^2 \bigg(\frac{H_\psi^2}{H^2}-\frac{H_{\psi\psi}}{H}\bigg).
\ear
  This expression for $W(u)$ directly generalizes those obtained previously in a number of 
  papers. Notably, \eqn{W} substantially simplifies if $H(\psi) = \const$ (i.e., if there is 
  only a canonical or phantom scalar without mutual transition)
  and/or $q_e =0$ (a pure magnetic field, in which case $P =0$ and $\delta F =0$).  
  
  The static nature of the background makes it possible to separate the variables assuming 
\beq                 \label{to_Omega}  
 			\df = \e^{i\Omega t} X(u),\qq \Omega = \const,
\eeq  
  which leads  to an ordinary differential equation for $X(u)$, suitable for further stability analysis,
\beq                \label{eq-X}
    	X'' - X' \bigg(\gamma'+ 2\beta'-\alpha'+ \frac{H'}{H}\bigg) 
    			+\Big[\e^{2\alpha-2\gamma}\Omega^2 - W(u)\Big] X =0,
\eeq    
  The next standard step is to reduce \eqn{eq-X} to a canonical form by invoking the ``tortoise''
  coordinate $z$ instead of $u$ and changing the unknown function $X(u) \mapsto Y(z)$ 
  according to
\beq                \label{u-z}
       \frac{du}{dz} = \e^{\gamma-\alpha}, \qq    
       X = Y \e^{-\eta}, \cm \eta' := \beta' + \frac{H'}{2H},
\eeq
   which results in the \Schr-like equation
\beq                                                        \label{Schr}
       Y_{zz} + [\Omega^2 - \Veff(z)] Y =0, \qq
        \Veff (z) = \e^{2\gamma-2\alpha} [W(u) + \eta''+ \eta'(\eta' + \gamma'-\alpha')], 
\eeq
  where $\Veff$ is the effective potential; the index $z$ denotes $d/dz$, while the prime,
  as before, stands for $d/du$. The potential $\Veff$ is expressed as a function of $u$, 
  and a transition to $x$ may be quite uneasy, involving solution of a transcendental equation. 

  If there is a nontrivial solution to (\ref{Schr}) with $\im \omega \leq 0$
  satisfying some physically reasonable conditions at the ends of the range of $z$ or 
  $u$ (in particular, the absence of ingoing waves), then the static system is unstable since 
  $\df$ can grow exponentially or linearly (if $\im \omega =0$) with $t$. 
  Otherwise our static system is stable in the linear approximation. Thus,  as usual in such 
  studies, the stability problem is reduced to a  boundary-value problem for \eq (\ref{Schr}) 
  --- see, e.g., \cite{kb-hod, kod03, stepan04, stepan05, gon08, stab11, stab12} and many 
  others. The \pb\ equations are equally valid in \ME\ and \MJ\ because \rf{map} is
  simply a substitution from the viewpoint of differential equations, and the conformal
  factor $f(\phi)$ is time-independent. However, the stability conclusions are not directly 
  extended to \MJ\ because the boundary conditions are properly formulated in
  the Jordan frame and can depend on the choice of STT and particular background solutions,
  as exemplified in \cite{we-1, we-2}.  
  
  Note that all the above relations are written without fixing the background
  radial coordinate $u$.
  
\section{NED-STT electrovacuum solutions}

 Exact analytical solutions of the theory \rf{S_J} (except for the Maxwell case $\cL(\cF) = \cF$) 
 are rather hard to obtain, even assuming $U(\phi) \equiv 0$ and even in such highly 
 symmetric and physically relevant situations as static spherical symmetry:
 to our knowledge, there are only a number of examples of numerical solutions in particular 
 STT and NED \cite{cle00, torii00, yaz1, yaz2, yaz3}, and some families of analytical solutions
 for electrically charged black holes with a Born-Infeld-dilaton source in GR have been 
 obtained in \cite{riazi06}.
 
 There is, however, a recently discovered circumstance that makes it possible to consider 
 many already known STT solutions with a Maxwell \emag\ field as those of NED-STT theory 
 within an arbitrary NED having a Lagrangian $\cL(\cF)$ with a correct Maxwell limit at 
 small $\cF$ and arbitrary STT: these are all solutions in which the \emag\ field satisfies 
 the condition $\cF \equiv 0$ \cite{dyon-we}.  
  
 Thus, in the case of spherical symmetry with the metric \rf{ds-E},
 the only nonzero components of $F\mn$ describe radial electric and magnetic fields with 
 the respective charges $q_e$ and $q_m$: in full similarity with \rf{F10} an \rf{F},
\beq                                                        \label{F_mn}
			F_{01} = - F_{10} = \frac{q_e}{\cL_\cF} \e^{\alpha - 2\beta +\gamma},
			\qq 
			F_{23} = - F_{32} = q_m \sin\theta, 
			\qq
			\cF = 2\e^{-4\beta}\bigg(q_m^2 -  \frac {q_e^2}{\cL_\cF^2}\bigg) ,
\eeq  
  where $\cL_\cF = \D\cL/\D\cF$. In particular, in Maxwell electrodynamics, with 
  $\cL\equiv \cF$, we have simply $\cF= 2(q_m^2 - q_e^2)\e^{-4\beta}$.
  Therefore, assuming $q_m^2 = q_e^2$, we obtain $F \equiv 0$. More than that, in 
  NED with a correct Maxwell limit, the condition $\cF = 0$ leads to $\cL_\cF =1$, which, 
  by \eqn{F_mn}, is preserved in the whole space-time along with $\cF \equiv 0$, and the 
  Maxwell \emag\ field is thus an exact solution of such a NED theory. More precisely
  \cite{dyon-we}, we can present $\cL$ as a Taylor expansion
\beq                    \label{Taylor}
  		\cL(\cF) = \cF + \sum_{k=2}^{\infty} a_k \cF^k, \cm a_k = \const.
\eeq  
  It means that if we assume $\cF \equiv 0$ in a solution to the Maxwell equations in 
  any given space-time, it is simultaneously a solution to NED equations since the
  latter contain only the first-order derivative of $\cL$ in $\cF$, this derivative is equal 
  to 1 at $\cF=0$, as in Maxwell's theory, and NED equations thus reduce to Maxwell's. 
  The corresponding stress-energy tensor also coincides with that of the Maxwell field.
   
  Returning to spherical symmetry, we can recall that the previously discussed \ssph\ 
  electrovacuum Maxwell-GR and Maxwell-STT solutions (e.g., \cite{penney, br73, we-2, cold-q}) 
  generally assumed only an electric nonzero charge since the very existence of magnetic 
  charges was subject to doubt. However, due to the electric-magnetic duality of Maxwell's 
  theory, in all such solutions one can safely replace $q^2 = q_e^2$ with 
  $q^2= q_e^2+ q_m^2$, and in addition we can assume $q_m^2 = q_e^2$,  which leads to 
  a zero \emag\ invariant $\cF$. In other words, we can use such solutions with 
  $q^2 = q_e^2 + q_m^2 = 2q_e^2 = 2q_m^2$ and assert that they satisfy the NED 
  equations coupled to those of GR or STT. 
  
  Our previous paper \cite{we-2} described in sufficient detail the known \ssph\ electrovacuum 
  solutions of GR \cite{penney} and a few known examples of STT 
  \cite{br73, BD-STT, barker, schwg, bruk94} along with their \pb\ dynamics. For completeness, 
  we here present the general form of these solutions, referring for details to \cite{we-2,br73} 
  and references therein. The metric in \MJ\ is
\beq 			\label{ds-sol}
		ds^2 = g\mn dx^\mu dx^\nu 
		= \frac{1}{f(\phi)}\bigg\{\frac{dt^2}{q^2\,s^2(h,u+u_1)} 
     		   - \frac{q^2\,s^2(h,u+u_1)}{s^2(k,u)}
	          	\bigg[\frac{du^2}{s^2(k,u)} + d\Omega^2\bigg]\bigg\},
\eeq   
   where $f(\phi)$ is the nonminimal coupling function of STT from \rf{S_J}, while
   the function of two arguments $s(a, x)$ is defined as follows:
\beq  			\label{def-s}
		s(a,x) \equiv \vars{ a^{-1}\sinh\,ax,  & a > 0; \\
     				       x, & a = 0;  \\
     			   a^{-1}\sin ax, & a < 0  }
\eeq     
  (one should substitute here $k$ or $h$ instead of $a$ and $u$ or $u+u_1$ instead of $x$).
  The \emag\ field is given by \eqs \rf{F_mn} with $\cL_\cF=1$, and, as said above, we now 
  interpret it as a NED solution assuming $q^2 = 2q_e^2 = 2q_m^2$. The scalar field $\psi$ 
  is simply $\psi = Cu$ and is related to $\phi$ according to \rf{phi-psi}; the quantities
  $k, h, C, u_1$ are integration constants satisfying the relations
\beq 				\label{constr}
		    k^2\sign k = \eps C^2 + h^2\sign h, \qq   s^2(h,\ u_1) = 1/q^2, 
\eeq   
  where the first equality follows from the constraint equation $G^1_1 = \ldots$ among the
  Einstein equations in \ME\ while the second one provides asymptotic flatness in \ME; 
  the factor $\eps = \pm 1$ is as introduced in \rf{phi-psi}.  
  
  The stability analysis in \cite{we-2} concerned the solutions \rf{ds-sol}--\rf{constr} with 
  a canonical scalar field, i.e., with $\eps =1$.  The solutions form a few families depending 
  on signs of the parameters $k$ and $h$ as well as on the properties of the function $f(\phi)$.
  Most of space-times belonging to this class possess naked singularities at the center 
  $r \equiv \e^\beta =0$, which can be scalar type (attractive for neutral test particles due to
  $g_{00} \to 0$) and \RN\ type (repulsive due to $g_{00}\to \infty$). Perturbations of these 
  solutions obey a common master equation \rf{Schr} with a certain effective potential, but 
  the relevant boundary conditions are formulated in \MJ\ and are different in different STT 
  due to diverse behaviors of the nonminimal coupling function  $f(\phi)$. In many cases, 
  as in Penney's solution, it turns out that linear physically relevant \pbs\ may grow at an 
  unbounded rate \cite{kb-hod}, which means that they should be analyzed in a 
  nonlinear regime from the very beginning despite their smallness. In other cases, there
  emerge boundary-value problems solved numerically to find out whether or not the 
  system is unstable, and stability ranges in the parameter spaces of the solutions have 
  been determined, see their description and tables in \cite{we-2}.     

\section{NED versus Maxwell in the solution stability conditions}

  Now let us find out whether our interpretation of the solutions \rf{ds-sol}--\rf{constr} as
  those of STT coupled to NED can change the stability conditions as compared to the same 
  solutions treated as electrovacuum STT-Maxwell ones. To do that, we should determine the
  possible changes in the effective potential $\Veff$ at a transition from Maxwell to NED. 
  Thus, in a general STT, we should write the potential in \rf{Schr} for $L(\psi, F)$ given 
  by \eqn{Lpsi}, that is, 
\beq                \label{Lpsi1}
		L(\psi, F) =  \frac 1{f^2(\phi(\psi))} \cL\big(f^2(\phi(\psi)) F\big),
\eeq  
  with the nonminimal coupling function $f(\phi)$ and $\phi(\psi)$ found from \rf{phi-psi}.
  Thus in \eqn{W} we should put $H(\psi) = \eps = \pm 1$ and find the following expressions 
  for the derivatives of $L$: 
\bearr                \label{L'}
		L_F = \cL', \qq  
		L_\psi = \frac{2f_\psi}{f^3}[f^2 F \cL' - \cL], \qq  
		L_{FF} = f^2 \cL'', \qq
		L_{F\psi} = 2 f f_\psi F \cL'', 
\nnn		
		L_{\psi\psi} = \frac{2f_{\psi\psi}}{f^3}[f^2 F \cL' - \cL]
					    + \frac{2f_\psi^2}{f^4}[2f^4 F^2 \cL'' - 3f^2 F \cL' -3 \cL],
\ear
   where the prime denotes a derivative of $\cL$ with respect to its argument. With these 
   expressions we can study the stability of any \ssph\ NED-STT solutions under monopole \pbs.
   
   Evidently, in the ``old'' case of Maxwell electrodynamics, $\cL = \cF$, we have $L_F=1$, while 
   all other derivatives in \rf{L'} are equal to zero, so that the expression \rf{W} simplifies to yield
\beq                  \label{W-Maxw}
		 W(u) = \frac{2 \eps \psi'^2}{\beta'^2} \e^{2\alpha-2\beta}
      					\big[(q_e^2+ q_m^2)\e^{-2\beta} -1 \big],
\eeq   
   in agreement with \cite{kb-hod} and \cite{we-2}. The symmetry between electric and 
   magnetic charges in \rf{W-Maxw} conforms to the electric-magnetic duality of Maxwell's
   theory. 

\paragraph{The case of Maxwell-like solutions with $q_e^2 = q_m^2$.}
   This case corresponds to $\cF = F =0$, and according to \eqn{Taylor}, it leads to $\cL =0$
   and $\cL' =1$. Thus a possible NED contribution to $W$ and hence to $\Veff$ can only 
   emerge due to $\cL'' \ne 0$. By \eqn{dF} we now have $P =0$, and it then follows that
   $W$ does not contain any NED contribution as compared to the Maxwell theory.
   
   Thus the linear stability properties of Maxwell-like NED-STT solutions with $\cF=0$ under 
   \sph\ \pbs\ entirely coincide with those of Maxwell-STT solutions analyzed in \cite{we-2}. 

\paragraph{The case of exceptional Brans-Dicke wormholes.}
  A case of interest among Maxwell-STT space-times is the exceptional \wh\ solution 
  \cite{we-25} in the Brans-Dicke (BD) STT \cite{BD-STT} characterized by $f(\phi) =\phi$ and 
  $H(\phi) = \omega/\phi$ in \eqn{S_J}, where $\omega$ is a coupling constant, so that 
  the $\phi$ field is canonical at $\omega > -3/2$ and phantom at $\omega < -3/2$. 
  The exceptional \wh\ solution is obtained at $\omega =0$ (thus in the canonical domain),
  so that in \eqn{phi-psi} and subsequent equations we have $\eps =1$ and 
  $f = \phi = \exp(2\psi/\sqrt{3})$. It corresponds to a special case of the solution 
  \rf{ds-sol}--\rf{constr} with the constants $k = 2h >0, C^2 = 3h^2$. In this case the limit 
  $u \to \infty$, corresponding to a singular center in \ME, leads to a regular sphere in \MJ,
  making it necessary to continue the solution beyond it by choosing a coordinate with a
  finite value at $u=\infty$. In \cite{we-25} it is chosen as $y = \e^{-2hu}$, so that $u = \infty$
  maps to $y=0$, and the solution is smoothly continued to negative $y$.  
  In terms of $y$ the metric in \MJ\ reads
\beq         \label{BD-wh}
	  ds_J^2 = \frac{4 h^2 dt^2}{[m+h - y(m-h)]^2} - \frac {4 [m+h - y(m-h)]^2}{(1-y^2)^2} 
	  		\bigg( \frac{4 dy^2}{(1-y^2)^2} + d\Omega^2 \bigg),
\eeq	
  where $m = \sqrt{h^2 + q^2}$, and the range of $y$ is $y \in (-1, 1)$, its two ends 
  corresponding to two flat spatial infinities of the \wh. The charge $q$ was interpreted as an 
  electric one in \cite{we-25}, but we can safely suppose, instead, a combination of electric 
  and magnetic charges, $q^2 = q_e^2 + q_m^2$. The \wh\ throat is located at  
\beq         \label{y_th}
           y = y_{\rm th} = \frac{\sqrt m - \sqrt h} {\sqrt m + \sqrt h},
\eeq       
  which is zero in the ``neutral'' case $q=0$ (when $m=h$), which makes the \wh\ symmetric
  relative to its throat, while at $q\ne 0$ it is asymmetric. This solution represents a special case
  of conformal continuation \cite{stepan04, kb-CC2}, when the whole manifold \ME\ maps to only a
  region of the manifold \MJ, namely, the region $y > 0$. The region $y < 0$ can be mapped 
  to another Einstein-frame manifold $\ME_-$, and it is identical to \ME\ only in the symmetric
  case $q=0$, in which the metric takes the especially simple form
\beq              \label{wh_0}
		ds_J^2\Big|_{q=0} = dt^2 - \frac{16 h^2}{(1-y^2)^2} 
								\bigg(\frac{4 dy^2}{(1-y^2)^2} + d\Omega^2\bigg)
				= dt^2 - \frac{d r_J^2}{1 - 4h/r_J}  - r_J^2 d\Omega^2,
\eeq      
  where $r_J = 4h/(1-y^2)$ is the spherical radius in \MJ.
  
  The stability study in \cite{we-25} showed that this solution is stable under \sph\ \pbs\ in the 
  general asymmetric case and unstable if $q=0$. The reason is that the stability analysis 
  implies solving two separate boundary-value problems, one in each Einstein-frame
  manifold, and viable \pb\ modes must unify solutions of these problem from the two
  manifolds since there should be a unified mode covering the whole \MJ. This is only possible 
  if these solutions provide coinciding eigenvalues $\Omega^2$, which takes place only in 
  the symmetric case where the two manifolds are identical. The absence of coinciding 
  eigenvalues at $q\ne 0$ means that, in particular, growing \pb\ modes are absent, i.e., 
  the \wh\ solution is stable. On the contrary, in the symmetric case $q=0$ the two 
  boundary-value problems yield growing modes with coinciding eigenvalues, indicating 
  instability of the \wh. 
  
  A natural question is whether or not these inferences will change if we invoke NED
  with $\cL(\cF)$ instead of Maxwell electrodynamics, assuming, as before, its correct 
  Maxwell limit at small $\cF$. 
  
  First of all, it is clear that the same exact solution as with the Maxwell field exists with 
  any such NED under the assumption that $\cF=0$ and $q_e^2 = q_m^2$, since then, 
  as we saw above, the \pb\ dynamics is quite the same as with the Maxwell field. 
    
  A more interesting opportunity can be imagined if $\cF\ne 0$ in the BD-Maxwell \wh\
  solution. First of all, it is not evident, under which choice of NED there exist similar \wh\ 
  solutions in the BD-NED framework: it is a separate problem to find such solutions.
  However, assuming that such a solution does exist, it is clear that when the charge $q$ 
  tends to zero (be it electric, or magnetic, or a combination of both), the limiting configuration
  will be the same ``neutral'' symmetric solution with the metric \rf{wh_0}. And in this case 
  one can inquire whether this solution, being unstable in the BD-Maxwell framework, 
  can become stable as a solution to BD-NED equations. This can happen if the effective 
  potential $\Veff$ of \pbs\ in the limit $q\to 0$ preserves some trace of the NED, which 
  can be different for $y>0$ from $y <0$. 
  
  However, a direct inspection of the general expression \rf{W} for $W(u)$ 
  shows that it is not the case. Indeed, at zero charge we have again $\cF=0$, $\cL=0$, and 
  $\cL_\cF=1$, and among the derivatives \rf{L'} the only nonzero expressions are $\cL_\cF$
  and possibly $\cL_{\cF\cF}$. However, these quantities enter in \rf{W} with the factor $P$, 
  which turns to zero at $q_e=0$. We conclude that with any NED under consideration, 
  the zero charge limit of \rf{W} yields the expression of $W(u)$ for a pure scalar, 
  in particular, if $H(\psi) = \eps = \pm 1$, we have 
  $W(u) = -  2 \eps \psi'^2 \e^{2\alpha-2\beta}/\beta'^2$.
  The same is true for the potential $\Veff$ given by \eqn{Schr} since the metrics of ``charged'' 
  solutions tend to those of ``neutral'' ones as $q\to 0$.

\section{Concluding remarks}

  We have obtained a general expression for the effective potential $\Veff$ of \sph\ \pbs\
  of \ssph\ scalar-electrovacuum solutions of GR with a sufficiently general
  scalar-\emag\ interaction given by the action \rf{S}. This action can be considered as
  that of the Einstein frame in the Bergman-Wagoner-Nordtvedt class of STT, therefore,
  it also enables the stability analysis of a wide choice of electrovacuum NED-STT solutions,
  both known and still unknown ones, with possible inclusion of arbitrary scalar self-interaction
  potentials $U(\phi)$, with canonical, phantom or partly phantom scalar fields.

  Our general observation made for the case of NED with a correct Maxwell weak-field limit 
  is that the zero charge limit of $\Veff$ does not contain any trace of the NED theory,
  therefore, the zero charge limit of any NED-STT solution exhibits the same \pb\ dynamics
  as the corresponding vacuum STT solution. This is true, in particular, for the previously
  studied exceptional Brans-Dicke \wh\ solution, proved to be stable at nonzero electric or
  magnetic charge (or both) and unstable without a charge.

  On the other hand, considering the existing family of NED-STT solutions with $q_e^2=q_m^2$
  and $\cF =0$, we have proved that their stability properties are quite the same as of their
  Maxwell-STT counterparts. Specifically, the condition $\cF = 0$ yields $\cL = 0$ and 
  $\cL_\cF = 1$, which reduces the stress-energy tensor of the nonlinear electromagnetic field 
  to that of the standard Maxwell theory. Consequently, the effective potential $\Veff$ 
  governing the monopole perturbations is identical to its Maxwellian counterpart, precluding 
  the need for any additional dynamic analysis. All stability theorems and analytical results 
  previously established for STT-Maxwell configurations are thus directly extended to this 
  specific class of STT-NED solutions without modification.
  
  In our forthcoming studies we hope to further consider the stability of existing and 
  possible new NED-STT solutions, both analytical and numerical ones. For stable ones, 
  in particular, \bh\ ones, a further analysis should involve quasinormal modes and other
  observational properties.
      
\Acknow 
     {The research of K. Bronnikov was supported in part by the Ministry of Science and 
      Higher Education of the Russian Federation, Project FSWU-2026-0010.
      R. Ibadov, F. Shaymanova and N. Shukurullokhon  gratefully acknowledge the support from
      Ministry of Innovative Development of the Republic of Uzbekistan, Project  
      No. FZ-20200929385. 
}   
        

\end{document}

%% file: preamble.tex
\usepackage{amsfonts,amssymb,cite}
\usepackage{graphicx}

\def\onecol{\onecolumn \mathindent 2em}

\def\noi{\noindent}

\newcommand{\Title}[1]{\noi {{\Large\bf #1}}\\[1ex]}

\def\Aunames#1{\noi{\bf #1}}
\def\au#1{${}^{#1}$}
\def\Addresses#1{\medskip\noi \protect
	\begin{description}\itemsep -3pt {\it #1} \end{description}}
\def\adr#1#2{\item[${}^{#1}$]{\it #2}}

\newcommand{\Abstract}[1]{\vskip 2mm \begin{center}
        \parbox{16.4cm}{\small\noi #1} \end{center}\medskip}

\def\email#1#2{\footnotetext[#1]{e-mail: #2}\addtocounter{footnote}{1}}

\def\nqq{\hspace*{-2em}}
\def\nhq{\hspace*{-0.5em}}

\def\qq{\qquad}
\def\cm{\hspace*{1cm}}
\def\inch{\hspace*{1in}}

\usepackage{color}

\def\Acknow#1{\subsection*{Acknowledgments} #1}

\def\Jl#1#2{#1 {\bf #2},\ }

\def\ApJ#1 {\Jl{Astroph. J.}{#1}}
\def\CQG#1 {\Jl{Class. Quantum Grav.}{#1}}
\def\DAN#1 {\Jl{Dokl. AN SSSR}{#1}}
\def\GC#1 {\Jl{Grav. Cosmol.}{#1}}
\def\GRG#1 {\Jl{Gen. Rel. Grav.}{#1}}
\def\IJMPD#1 {\Jl{Int. J. Mod. Phys. D}{#1}}
\def\JETF#1 {\Jl{Zh. Eksp. Teor. Fiz.}{#1}}
\def\JETP#1 {\Jl{Sov. Phys. JETP}{#1}}
\def\JHEP#1 {\Jl{JHEP}{#1}}
\def\JMP#1 {\Jl{J. Math. Phys.}{#1}}
\def\NPB#1 {\Jl{Nucl. Phys. B}{#1}}
\def\NP#1 {\Jl{Nucl. Phys.}{#1}}
\def\PLA#1 {\Jl{Phys. Lett. A}{#1}}
\def\PLB#1 {\Jl{Phys. Lett. B}{#1}}
\def\PRD#1 {\Jl{Phys. Rev. D}{#1}}
\def\PRL#1 {\Jl{Phys. Rev. Lett.}{#1}}

\def\al{&\nhq}
\def\lal{&&\nqq {}}
\def\eq{Eq.\,}
\def\eqs{Eqs.\,}
\def\beq{\begin{equation}}
\def\eeq{\end{equation}}
\def\bear{\begin{eqnarray}}
\def\bearr{\begin{eqnarray} \lal}
\def\ear{\end{eqnarray}}
\def\earn{\nonumber \end{eqnarray}}

\def\nnn{\nonumber\\ \lal }

\def\yy{\\[5pt] {}}
\def\yyy{\\[5pt] \lal }
\def\eql{\al =\al}

\def\dst{\displaystyle}
\def\tst{\textstyle}
\def\fracd#1#2{{\dst\frac{#1}{#2}}}
\def\fract#1#2{{\tst\frac{#1}{#2}}}
\def\Half{{\fracd{1}{2}}}
\def\half{{\fract{1}{2}}}

\def\e{{\,\rm e}}
\def\D{\partial}

\def\im{\mathop{\rm Im}\nolimits}

\def\sign{\mathop{\rm sign}\nolimits}

\def\const{{\rm const}}
\def\eps{\varepsilon}

\newcommand{\vars}[1]{\left\{\begin{array}{ll}#1\end{array}\right.}

